\documentclass[a4paper]{article}
\usepackage{amssymb}
\usepackage{amsmath}
\usepackage{latexsym}
\usepackage{mathrsfs}
\usepackage{cancel}
\usepackage{hyperref}
\usepackage{amsrefs}

\providecommand{\keywords}[1]{Keywords : #1}

\begin{document}

\title{Exact form of Maxwell's equations and Dirac's magnetic monopole in Fock's nonlinear relativity }
\author{N. Takka\hspace{0.1cm}\footnote{E-mail: takka.naimi@gmail.com} \hspace{0.1cm}and\hspace{0.1cm} A. Bouda\hspace{0.1cm}\footnote{E-mail: bouda{\_} a@yahoo.fr}\hspace{0.1cm}\\  Laboratoire de Physique Théorique, Faculté des Sciences Exactes,\\
Université de Bejaia, 06000 Bejaia, Algeria}
\date{\today}

\maketitle

\begin{abstract}
After having obtained previously an extended first approximation of Maxwell's equations in Fock's nonlinear relativity, we propose here the corresponding exact form. In order to achieve this goal, we were inspired mainly by the special relativistic version of Feynman's proof from which we constructed a formal approach more adapted to the noncommutative algebra. This reasoning lets us establish the exact form of the generalized first group of Maxwell's equations. To deduce the second one, we have imposed the electric-magnetic duality. As in the $k$-Minkowski space-time, the generalized Lorentz force depends on the mass of the particle. After having restored the $R$-Lorentz algebra symmetry, we have used the perturbative treatment to find the exact form of
the generalized Dirac's magnetic monopole in our context. As consequence, the Universe could locally contain the magnetic charge but in its totality it is  still neutral.
\end{abstract}

\keywords{Fock-Lorentz transformation, $R$-Minkowski spacetime, Maxwell's equations, magnetic monopole.}

\newpage


\section{Introduction}
At the end of the last century, Dyson published Feynman's proof of the homogeneous Maxwell's equations in the context of the noncommutative formalism through the use of the commutation relations between position and velocity for a nonrelativistic particle \cite{Dyson}. According to Dyson, Feynman’s aim was to go as far as possible beyond the framework of the conventional physics by making as few assumptions as he could. Trying to explore this new area  of research, Feynman was aware that the discovery of the new physical laws involves the non-use of certain fundamental ideas and then he avoided assuming momentum. However, after the calculation, he realized that he did not discover a new theory but just another way equivalent to the standard approach. Some years later, Tanimura formulated both a special relativistic and a general relativistic versions of this derivation \cite{Tanimura}. Later Bérard and al. extended this approach in order to derive the two groups of Maxwell's equations \cite{Berard1} and then they studied in many works, the emergence of the magnetic monopole as a consequence of the restoration of the Lorentz algebra symmetry,
e.g. \cite{Berard2} and \cite{Berard3}. Recently, Harikumar and al. established Maxwell's equations valid to the first order in the $k$-Minkowski spacetime \cite{Harikumar1} and \cite{Harikumar2}. In \cite{Harikumar2} and \cite{Montesinos1}, the equivalence between the minimal coupling prescription and Feynman’s approach has been discussed in the noncommutative space-time  and in the relativistic classical mechanics, respectively. Finally, to first order in the deformation parameter, we have proposed in \cite{Takka1} an extended form of Maxwell's equations in the context of the Fock-Lorentz transformation. \\

In this work, we started from the equivalence between Feynman's proof of Maxwell's equations and the standard proof in the framework of the lagrangian formalism, pointed out since the first paper \cite{Dyson} in order to provide a new approach within which the electromagnetism can be studied formally. In this perspective, the generalization of electromagnetism in any general theory than special relativity can be envisaged by the exclusive knowledge of the commutation relations $[x^{\mu}, x^{\nu}]$, $[x^{\mu}, p^{\nu}] $, $[p^{\mu}, p^{\nu}]$ and the explicit form of the four-dimensional momentum $p^{\mu}$ valid in the absence of the electromagnetic field $F^{\mu\nu}$. To this end, we have exploited two advantages related to the use of the noncommutative formalism, namely, on the one hand, the manifestation of the electromagnetic field by the noncommutativity of velocities in Feynman's approach, and on the other hand, the development of the latest version of Fock's nonlinear relativity in \cite{Bouda-Foughali}. \\
 
Beside Deformed Special Relativity, also known as Doubly Special Relativity (DSR) which keeps invariant, the speed of light and the Planck energy $E_{p}$ or the Planck length $l_{p}$ \cite{Ghosh}-\cite{Magueijo2}, Fock's nonlinear relativity constitutes another audacious attempt to go beyond the conventional physics, by exploring the implications of the non-constancy of the speed of light in vacuum for different inertial frames. This curiosity is mainly motivated by the desire to resolve certain problems and paradoxes in cosmology \cite{Albrecht-Magueijo} and in quantum gravity \cite{Magueijo1}. 
Mathematically, its development has undergone several steps to obtain a coherent theory. To summarize briefly this evolution, let us take as origin \cite{Fock} in which it has been proposed that from the first principle of relativity alone, the general form of the coordinates transformation between inertial frames could be written as

\begin{equation}\label{equation1}
t^{\prime }=\frac{\gamma\big(t-ux/c^2\big)}{\alpha_R},\quad x^{\prime }=\frac{
\gamma(x-ut)}{\alpha_R},\quad y^{\prime }=
                                         \frac{y}{\alpha_R},\quad z^{\prime}
                                                                           =\frac{z}{\alpha_R},
\end{equation}
here obviously $\gamma =\big(1-u^{2}/c^{2}\big)^{-1/2}$ and ${\alpha _{R}}=1+\left[(\gamma -1)ct-\gamma xu/c\right]/R$ where $R$ designates the radius of the Universe. By inspiring from \cite{Ghosh} in which the coordinates transformation of DSR \cite{Amelino1, Amelino2, Magueijo1, Magueijo2} has been recovered, recently, they have succeeded in \cite{Bouda-Foughali} to reproduce the Fock-Lorentz transformations and to suggest the momenta transformation after having defined the appropriate deformed Poisson brackets. The main novelty of this rewriting is that the contraction $x_{\mu}p^{\mu}$ becomes an invariant which made possible the coherent description of free particles by means of plane waves. In \cite{Foughali-Bouda1} and \cite{Foughali-Bouda2}, the correspondence between the $R$-Minkowski spacetime and de Sitter spacetime has been established after having derived the Klein-Gordon and Dirac equations, respectively. On a quest for new generalizations even more satisfying, we have chosen once more to focus on Maxwell's equations and Dirac's magnetic monopole in the context of Fock's nonlinear relativity. For this purpose, we propose in what follows to go further than the previous work valid only to first order in the deformation parameter \cite{Takka1}, by establishing the corresponding exact formulation.\\ 

This paper is organized as follows. In section 2, we have developed the last version of Feynman's proof which allowed us to derive the  exact form of the generalized Maxwell's equations.  In section 3, we have highlighted the link between the restoration of the $R$-Lorentz algebra symmetry and the emergence of the magnetic monopole. By using the perturbative treatment, we have obtained the exact form of the generalized Dirac's magnetic monopole in our context. After that, by calculating the flux of the magnetic field through  a closed spherical surface, we have shown that the Universe could locally contain the magnetic charge but in its totality it is still neutral. Section 4 is devoted to conclusion.

\section{Maxwell's equations}
From \cite{Foughali-Bouda1} and \cite{Foughali-Bouda2}, the quantization of the $R$-deformed phase space algebra gives us the following result

\begin{align}
\label{equation2}
\left[x^\mu,x^\nu\right] & = 0, \\
\label{equation2a}
\left[x^\mu,p^\nu\right] & = - i \hbar \eta^{\mu\nu} + \frac{i \hbar}{R}\eta^{0\nu}x^\mu, \\
\label{equation2b}
\left[p^\mu,p^\nu\right] & = -\frac{i \hbar}{R}\left(p^\mu\eta^{0\nu}-p^\nu\eta^{\mu0}\right),
\end{align}
where $\eta^{\mu\nu} = (+1,-1,-1,-1)$ and $\mu ,\nu =0,1,2,3$. Now, let us introduce the symmetrization operator symbolized by $<...>$ \cite{Tanimura} in such a way that the four-dimensional momentum valid in the absence of the electromagnetic field \cite{Bouda-Foughali} becomes

\begin{equation}\label{equation3}
p^{\mu}=m\Big<\Big(1-\frac{x^{0}}{R}\Big)^{-2}\dot{x}^{\mu}\Big>+\frac{m}{R}\Big<\Big(1-\frac{x^{0}}{R}\Big)^{-3}x^{\mu}\dot{x}^{0}\Big>.
\end{equation}
Given that the generalized Feynman's proof is conditioned by the knowledge of $[x^{\mu}, x^{\nu}]$, $[x^{\mu}, \dot{x}^{\nu}]$ and $[\dot{x}^{\mu}, \dot{x}^{\nu}]$, it is clear that we need to find the two last commutators. First, without doing any calculation, combining together the commutativity of the coordinates (\ref{equation2}), the linearity of Eq. (\ref{equation3}) with respect to the velocity and by observing the right side of Eq. (\ref{equation2a}), we can deduce that $[x^{\nu}, \dot{x}^{\lambda}]$ does not depend on the velocity

\begin{equation}\label{commutatif}
[x^{\mu}, [x^{\nu}, \dot{x}^{\lambda}]]=0,
\end{equation}
 and then 

\begin{eqnarray}\label{equation4}
\left[x^{\mu}, p^{\nu}\right]\hspace{-0.2cm}&=&\hspace{-0.2cm}\left[x^{\mu}, m\Big<\Big(1-\frac{x^{0}}{R}\Big)^{-2}\dot{x}^{\nu}\Big>+\frac{m}{R}\Big<\Big(1-\frac{x^{0}}{R}\Big)^{-3}x^{\nu}\dot{x}^{0}\Big>\right] \nonumber\\
                  \hspace{-0.2cm}&=&\hspace{-0.2cm}\left[x^{\mu}, m\Big(1-\frac{x^{0}}{R}\Big)^{-2}\dot{x}^{\nu}+\frac{m}{R}\Big(1-\frac{x^{0}}{R}\Big)^{-3}x^{\nu}\dot{x}^{0}\right] \nonumber\\
                  \hspace{-0.2cm}&=&\hspace{-0.2cm}m\Big(1-\frac{x^{0}}{R}\Big)^{-2}[x^{\mu}, \dot{x}^{\nu}]+\frac{m}{R}\Big(1-\frac{x^{0}}{R}\Big)^{-3}x^{\nu}[x^{\mu},\dot{x}^{0}].
\end{eqnarray}
By identifying Eq. (\ref{equation4}) with the right side of Eq. (\ref{equation2a}), we get

\begin{equation}\label{equation5}
[x^{\mu}, \dot{x}^{\nu}]=\frac{i\hbar}{m}\Big(1-\frac{x^{0}}{R}\Big)^{2}\Big(-\eta^{\mu\nu}+\frac{1}{R}\eta^{0\nu}x^{\mu}\Big)
-\frac{1}{R}\Big(1-\frac{x^{0}}{R}\Big)^{-1}x^{\nu}[x^{\mu}, \dot{x}^{0}],
\end{equation}
which, by taking $\nu=0$, gives
 
 \begin{equation}\label{equation6}
[x^{\mu}, \dot{x}^{0}]=\frac{i\hbar}{m}\Big(1-\frac{x^{0}}{R}\Big)^{3}\Big(-\eta^{\mu 0}+\frac{x^{\mu}}{R}\Big).
\end{equation}
After the substitution of Eq. (\ref{equation6}) in (\ref{equation5}), we obtain 

\begin{equation}\label{equation7}
[x^{\mu}, \dot{x}^{\nu}]=\frac{i\hbar}{m}\Big(1-\frac{x^{0}}{R}\Big)^{2}\left\{-\eta^{\mu\nu}+\frac{1}{R}\big(\eta^{0\nu}x^{\mu}+\eta^{\mu 0}x^{\nu}\big)-\frac{x^{\mu}x^{\nu}}{R^{2}}\right\}.
\end{equation}
In the first-order approximation, we can easily prove that Eq. (\ref{equation7}) reproduces the result obtained in \cite{Takka1}. We also remark that 

\begin{equation}\label{equationaj1}
[x^{\mu}, \dot{x}^{\nu}]=[x^{\nu}, \dot{x}^{\mu}].
\end{equation}
At this stage, we must look for the commutator involving only the velocities. For this purpose, we first need to explicit expression (\ref{equation3}) of the four-momentum. At the beginning, we start from

\begin{align}\label{equation8}
\dot{x}^{\mu}x^{0}&=x^{0}\dot{x}^{\mu}-[x^{0}, \dot{x}^{\mu}].
\end{align}
From this last result and taking into account Eq. (\ref{commutatif}), it is easy to show that for any natural number $n\in\mathbb{N}$,

\begin{equation}\label{equation9}
\dot{x}^{\mu}(x^{0})^{n}=(x^{0})^{n}\dot{x}^{\mu}-n(x^{0})^{n-1}[x^{0}, \dot{x}^{\mu}],
\end{equation}
and then

\begin{eqnarray}\label{equation13}
\big<(x^{0})^{n}\dot{x}^{\mu}\big>\hspace{-0.2cm}&=&\hspace{-0.2cm}\frac{1}{(n+1)}\Big\{(x^{0})^{n}\dot{x}^{\mu}+(x^{0})^{n-1}\dot{x}^{\mu}x^{0}\nonumber\\
&&+...+x^{0}\dot{x}^{\mu}(x^{0})^{n-1}+\dot{x}^{\mu}(x^{0})^{n}\Big\}\nonumber\\
                          \hspace{-0.2cm}&=&\hspace{-0.2cm}(x^{0})^{n}\dot{x}^{\mu}-\frac{n}{2}(x^{0})^{n-1}[x^{0}, \dot{x}^{\mu}]\nonumber\\
                          \hspace{-0.2cm}&=&\hspace{-0.2cm}(x^{0})^{n}\dot{x}^{\mu}-\frac{1}{2}\frac{d(x^{0})^{n}}{dx^{0}}[x^{0}, \dot{x}^{\mu}].
\end{eqnarray}
From this last result, by developing any regular function $f$ of $x^{0}$ in a power series, we straightforwardly get the following generalization

\begin{equation}\label{equation13r}
\big<f(x^{0})\dot{x}^{\mu}\big>=f(x^{0})\dot{x}^{\mu}-\frac{1}{2}\frac{d f(x^{0})}{dx^{0}}[x^{0}, \dot{x}^{\mu}].
\end{equation}
In order to determine the second term of expression (\ref{equation3}), let us notice that for three operators $A$, $B$, and $C$ satisfying the following commutation relations 

\begin{equation}\label{equationsym2}
 [A,B]=0\hspace{0.2cm}\text{and}\hspace{0.2cm}[A,[B,C]]=0,
\end{equation}
we have

\begin{equation}\label{equationsym6}
\big<ABC\big>=\frac{1}{2}A\big<BC\big>+\frac{1}{2}\big<BC\big>A.
\end{equation}
In fact,

\begin{eqnarray}\label{equationsym3}
\big<ABC\big>\hspace{-0.2cm}&=&\hspace{-0.2cm}\frac{1}{6}\Big[ABC+ACB+BAC+BCA+CAB+CBA\Big]\nonumber\\
\hspace{-0.2cm}&=&\hspace{-0.2cm}\frac{1}{6}\Big[A\big(BC+CB\big)+\big(BC+CB\big)A+BAC+CAB\Big].
\end{eqnarray}
Using Eq. (\ref{equationsym2}), we can easily show that the two last terms in (\ref{equationsym3}) can be written as follows

\begin{equation}\label{equationsym5}
BAC+CAB=\frac{1}{2}A\big(BC+CB\big)+\frac{1}{2}\big(BC+CB\big)A.
\end{equation}
If we substitute the last expression in (\ref{equationsym3}), we straightforwardly obtain Eq. (\ref{equationsym6}). By means of (\ref{equationsym6}), the momentum  written in (\ref{equation3}) can be expressed as

\begin{multline}\label{equationsym7}
p^{\mu}=m\Big<\Big(1-\frac{x^{0}}{R}\Big)^{-2}\dot{x}^{\mu}\Big>+\frac{m}{2R}\Big[x^{\mu}\Big<\Big(1-\frac{x^{0}}{R}\Big)^{-3}\dot{x}^{0}\Big>\\
       +\Big<\Big(1-\frac{x^{0}}{R}\Big)^{-3}\dot{x}^{0}\Big>x^{\mu}\Big].
\end{multline}
Finally, via the use of Eqs. (\ref{equation2}), (\ref{equation7}), (\ref{equationaj1}) and (\ref{equation13r}), we can check that (\ref{equationsym7}) reduces to the following expression

\begin{equation}\label{equation15}
p^{\mu}=m\Big(1-\frac{x^{0}}{R}\Big)^{-3}\left\{\dot{x}^{\mu}+\frac{1}{R}\big(x^{\mu}\dot{x}^{0}-x^{0}\dot{x}^{\mu}\big)\right\}+\frac{3i\hbar}{2R}\eta^{\mu 0}.
\end{equation}
To continue, let us return once more to (\ref{commutatif}) to be able to write

\begin{eqnarray}\label{equation19}
\left[\Big(1-\frac{x^{0}}{R}\Big)^{-n}, \dot{x}^{\mu}\right]\hspace{-0.2cm}&=&\hspace{-0.2cm}\frac{d}{dx^{\alpha}}\left\{\Big(1-\frac{x^{0}}{R}\Big)^{-n}\right\}[x^{\alpha}, \dot{x}^{\mu}]\nonumber\\
\hspace{-0.2cm}&=&\hspace{-0.2cm}\frac{d}{dx^{0}}\left\{\Big(1-\frac{x^{0}}{R}\Big)^{-n}\right\}[x^{0}, \dot{x}^{\mu}]. 
\end{eqnarray}
By making use of (\ref{equation15}) in the commutator $[p^{\mu}, p^{\nu}]$ and taking into account relations
(\ref{equation2}), (\ref{equation7}) and (\ref{equation19}), after a tedious calculation, we obtain

\begin{eqnarray}\label{equation20}
[p^{\mu}, p^{\nu}]\hspace{-0.2cm}&=&\hspace{-0.2cm} m^{2}\Big(1-\frac{x^{0}}{R}\Big)^{-4}[\dot{x}^{\mu}, \dot{x}^{\nu}]+\frac{m^{2}}{R}\big(1-\frac{x^{0}}{R}\big)^{-5}\Big\{x^{\nu}[\dot{x}^{\mu}, \dot{x}^{0}]\nonumber\\
&&+x^{\mu}[\dot{x}^{0}, \dot{x}^{\nu}]\Big\}+\frac{2i\hbar m}{R}\Big(1-\frac{x^{0}}{R}\Big)^{-2}\big(\eta^{\mu 0}\dot{x}^{\nu}-\eta^{\nu 0}\dot{x}^{\mu}\big)\nonumber\\
&&+\frac{2i\hbar m}{R^{2}}\Big(1-\frac{x^{0}}{R}\Big)^{-3}\big(\eta^{\mu 0}x^{\nu}-\eta^{\nu 0}x^{\mu}\big)\dot{x}^{0}.
\end{eqnarray}
Furthermore, if we replace Eq. (\ref{equation15}) in the right side of (\ref{equation2b}), we find

\begin{multline}\label{equation21}
[p^{\mu}, p^{\nu}]=-\frac{i\hbar m}{R}\Big(1-\frac{x^{0}}{R}\Big)^{-2}\big(\eta^{\nu 0}\dot{x}^{\mu}-\eta^{\mu 0}\dot{x}^{\nu}\big)\\
-\frac{i\hbar m}{R^{2}}\Big(1-\frac{x^{0}}{R}\Big)^{-3}\big(\eta^{\nu 0}x^{\mu}-\eta^{\mu 0}x^{\nu}\big)\dot{x}^{0}.
\end{multline}
After the identification, relations (\ref{equation20}) and (\ref{equation21}) give

\begin{eqnarray}\label{equation21a}
[\dot{x}^{\mu}, \dot{x}^{\nu}]\hspace{-0.2cm}&=&\hspace{-0.2cm}-\frac{1}{R}\Big(1-\frac{x^{0}}{R}\Big)^{-1}\left\{x^{\nu}[\dot{x}^{\mu}, \dot{x}^{0}]+x^{\mu}[\dot{x}^{0}, \dot{x}^{\nu}]\right\}\nonumber\\
&&-\frac{i\hbar}{mR}\Big(1-\frac{x^{0}}{R}\Big)^{2}\left\{\big(\eta^{\mu 0}\dot{x}^{\nu}-\eta^{\nu 0}\dot{x}^{\mu}\big)\right\}\nonumber\\
&&-\frac{i\hbar}{mR^{2}}\Big(1-\frac{x^{0}}{R}\Big)\left\{\big(\eta^{\mu 0}x^{\nu}-\eta^{\nu 0}x^{\mu}\big)\right\}\dot{x}^{0},
\end{eqnarray}
which, by taking $\nu=0$ and $\mu=0$, yields respectively 

\begin{multline}\label{equation21aa}
[\dot{x}^{\mu}, \dot{x}^{0}]=-\frac{i\hbar}{mR}\Big(1-\frac{x^{0}}{R}\Big)^{3}\left(\eta^{\mu 0}\dot{x}^{0}-\dot{x}^{\mu}\right)\\
-\frac{i\hbar}{mR^{2}}\Big(1-\frac{x^{0}}{R}\Big)^{2}\left(\eta^{\mu 0}x^{0}-x^{\mu}\right)\dot{x}^{0},
\end{multline}
and

\begin{multline}\label{equation21ab}
[\dot{x}^{0}, \dot{x}^{\nu}]=-\frac{i\hbar}{mR}\Big(1-\frac{x^{0}}{R}\Big)^{3}\left(\dot{x}^{\nu}-\eta^{\nu 0}\dot{x}^{0}\right)\\
-\frac{i\hbar}{mR^{2}}\Big(1-\frac{x^{0}}{R}\Big)^{2}\left(x^{\nu}-\eta^{\nu 0}x^{0}\right)\dot{x}^{0}.
\end{multline}
Finally, if we replace the two commutators (\ref{equation21aa}) and (\ref{equation21ab}) in the right side of Eq. (\ref{equation21a}), we get

\begin{equation}\label{equation22}
[\dot{x}^{\mu}, \dot{x}^{\nu}]=-\frac{i\hbar}{mR}\Big(1-\frac{x^{0}}{R}\Big)^{2}\left\{\big(\eta^{\mu 0}\dot{x}^{\nu}-\eta^{0\nu}\dot{x}^{\mu}\big)
-\frac{1}{R}\big(x^{\mu}\dot{x}^{\nu}-x^{\nu}\dot{x}^{\mu}\big)\right\}.
\end{equation}
To first approximation in $1/R$, expression (\ref{equation22}) reproduces the result found in \cite{Takka1}. Up to now, we have not included the electromagnetic field in the calculation. To do it, let us recall the existence of the canonical variables found in \cite{Bouda-Foughali} 

\begin{equation}\label{equation23}
X^{\mu}=\Big(1-\frac{x^0}{R}\Big)^{-1}x^{\mu}, \hspace{0.5cm}P^{\mu}=\Big(1-\frac{x^0}{R}\Big)p^{\mu},
\end{equation}
obeying the usual commutators of the special relativity \cite{Takka1}

\begin{equation}\label{equation24}
[ X^{\mu}, X^{\nu}]=0,\hspace{0.5cm} [X^{\mu}, P^{\nu}]= -i \hbar \eta^{\mu\nu},\hspace{0.5cm} [ P^{\mu}, P^{\nu}]=0.
\end{equation}
Knowing that in the presence of electromagnetic field

\begin{equation}
P^{\mu}=m\dot{X}^{\mu}+\frac{q}{c}A^{\mu}(X),
\end{equation}
hence, it results

\begin{equation}\label{equation25}
[X^{\mu}, \dot{X}^{\nu}]=-\frac{i \hbar}{m} \eta^{\mu\nu}.
\end{equation}
Because $[X^{\mu}, \dot{X}^{\nu}]$ has the same expression whether in presence or in absence of electromagnetic field, expression (\ref{equation7}) keeps its form unchanged even in the presence of this latter. To do the same with Eq. (\ref{equation22}), we introduce the generalized electromagnetic part in order to have

\begin{multline}\label{equation26}
[\dot{x}^{\mu}, \dot{x}^{\nu}]=-\frac{i\hbar q}{m^{2}}F^{\mu\nu}-\frac{i\hbar}{mR}\Big(1-\frac{x^{0}}{R}\Big)^{2}\Big\{\big(\eta^{\mu 0}\dot{x}^{\nu}-\eta^{0\nu}\dot{x}^{\mu}\big)\\
-\frac{1}{R}\big(x^{\mu}\dot{x}^{\nu}-x^{\nu}\dot{x}^{\mu}\big)\Big\},
\end{multline}
which is equivalent to

\begin{multline}\label{equation27}
F^{\mu\nu}=-\frac{m^{2}}{i\hbar q}[\dot{x}^{\mu}, \dot{x}^{\nu}]-\frac{m}{qR}\Big(1-\frac{x^{0}}{R}\Big)^{2}\Big\{\big(\eta^{\mu 0}\dot{x}^{\nu}-\eta^{0\nu}\dot{x}^{\mu}\big)\\
-\frac{1}{R}\big(x^{\mu}\dot{x}^{\nu}-x^{\nu}\dot{x}^{\mu}\big)\Big\}.
\end{multline}
Here of course, $F^{\mu\nu}$ is an antisymmetric tensor which can be, in general, a function of coordinates and velocities. However, as in \cite{Takka1}, we can check that $F^{\mu\nu}$ depends only on position. Indeed, after a long calculation, by using the Jacobi identity involving one coordinate and two velocities where

\begin{equation}
[x^{\lambda}, [\dot{x}^{\mu}, \dot{x}^{\nu}]]=-[\dot{x}^{\nu}, [x^{\lambda}, \dot{x}^{\mu}]]-[\dot{x}^{\mu}, [\dot{x}^{\nu}, x^{\lambda}]],
\end{equation}
we can make sure that Eqs. (\ref{equation2}), (\ref{equation7}) and (\ref{equation26}) allow to find

\begin{eqnarray}
\big[x^{\lambda}, F^{\mu\nu}\big]
\hspace{-0.2cm}&=&\hspace{-0.2cm}-\frac{m^{2}}{i\hbar q}[x^{\lambda}, [\dot{x}^{\mu}, \dot{x}^{\nu}]]-\frac{m}{qR}\Big(1-\frac{x^{0}}{R}\Big)^{2}\Big\{\eta^{\mu 0}[x^{\lambda}, \dot{x}^{\nu}]\nonumber\\
&&\hspace{1.2cm}-\eta^{0\nu}[x^{\lambda}, \dot{x}^{\mu}]-\frac{x^{\mu}}{R}[x^{\lambda}, \dot{x}^{\nu}]+\frac{x^{\nu}}{R}[x^{\lambda}, \dot{x}^{\mu}] \Big\}\nonumber\\
                         \hspace{-0.2cm}&=&\hspace{-0.2cm}0,
\end{eqnarray}
which means that $F^{\mu\nu}=F^{\mu\nu}(x)$. At this stage, we are able to derive the first group of the generalized Maxwell's equations in our case. In fact, let us  proceed as in \cite{Tanimura} by starting from the following Jacobi identity

\begin{equation}\label{equation28}
[\dot{x}^{\mu}, [\dot{x}^{\nu}, \dot{x}^{\lambda}]]+[ \dot{x}^{\lambda}, [\dot{x}^{\mu} , \dot{x}^{\nu}]]
 + [\dot{x}^{\nu}, [\dot{x}^{\lambda} , \dot{x}^{\mu}]] = 0.
\end{equation}
Substituting Eq. (\ref{equation26}) in (\ref{equation28}) and taking into account Eq. (\ref{equation2}) and (\ref{equation7}), we get after a long calculation

\begin{eqnarray}\label{equation29}
&&\hspace{-1.3cm}\big[\dot{x}^{\mu}, F^{\nu\lambda}\big]+\big[ \dot{x}^{\lambda}, F^{\mu\nu}\big]
+\big[\dot{x}^{\nu},  F^{\lambda\mu}\big]+\frac{2m}{qR}\Big(1-\frac{x^{0}}{R}\Big)^{2}\Big\{\Big(\eta^{0\nu}-\frac{x^{\nu}}{R}\Big)[\dot{x}^{\mu}, \dot{x}^{\lambda}]\nonumber\\
 &&\hspace{-1cm}-\Big(\eta^{0\lambda}-\frac{x^{\lambda}}{R}\Big)[\dot{x}^{\mu}, \dot{x}^{\nu}]+\Big(\eta^{0\mu}-\frac{x^{\mu}}{R}\Big)[\dot{x}^{\lambda}, \dot{x}^{\nu}]\Big\}=0.
\end{eqnarray}
Making the use of Eq. (\ref{commutatif}), it follows that

\begin{equation}\label{equation30}
[\dot{x}^{\mu}, F^{\nu\lambda}] = [\dot{x}^{\mu}, x^{\alpha}]\frac{\partial F^{\nu\lambda}}{\partial x^{\alpha}}.
\end{equation}
Finally, the use of Eqs. (\ref{equation2}), (\ref{equation7}), (\ref{equation26}) and (\ref{equation30}) lets us develop (\ref{equation29}), to find after a long calculation, the exact form of the following generalized first group of Maxwell's equations 

\begin{eqnarray}\label{equation31}
&&\big(\partial^{\lambda}F^{\mu\nu}+ \partial^{\nu}F^{\lambda\mu}+\partial^{\mu}F^{\nu\lambda}\big)-\frac{1}{R}\Big[x^{\alpha}\big(\eta^{0\mu}\partial_{\alpha}F^{\nu\lambda}+\eta^{0\lambda}\partial_{\alpha}F^{\mu\nu}\nonumber\\
&&+\eta^{0\nu}\partial_{\alpha}F^{\lambda\mu}\big)+\big(x^{\mu}\partial_{0}F^{\nu\lambda}+x^{\lambda}\partial_{0}F^{\mu\nu}
+x^{\nu}\partial_{0}F^{\lambda\mu}\big)-2\big(\eta^{0\lambda}F^{\mu\nu}\nonumber\\
&&+\eta^{0\nu}F^{\lambda\mu}+\eta^{0\mu}F^{\nu\lambda}\big)\Big]+\frac{1}{R^{2}}\Big[x^{\alpha}\big(x^{\mu}\partial_{\alpha}F^{\nu\lambda}+x^{\lambda}\partial_{\alpha}F^{\mu\nu}\nonumber\\
&&+x^{\nu}\partial_{\alpha}F^{\lambda\mu}\big)-2\big(x^{\lambda}F^{\mu\nu}+x^{\nu}F^{\lambda\mu}+x^{\mu}F^{\nu\lambda}\big)\Big]=0.
\end{eqnarray}
As a first remark, we can check that to first order in $1/R$, the last expression recovers the result established in \cite{Takka1}. In order to seek the second group, we proceed as in \cite{Takka1}. At the beginning, we contract expression (\ref{equation31}) by means of the metric tensor $\eta_{\mu\nu}$ to get

\begin{eqnarray}\label{equation32}
&&\hspace{-1.2cm}\big(\partial_{\mu}F^{\mu\lambda}+\partial_{\mu}F^{\lambda\mu}\big)-\frac{1}{R}\Big[x^{\mu}\big(\partial_{\mu}F^{0\lambda}
+\partial_{\mu}F^{\lambda 0}\big)+x_{\mu}\big(\partial_{0}F^{\mu\lambda}+\partial_{0}F^{\lambda\mu}\big)\nonumber\\
&&\hspace{-1.2cm}-2\big(F^{0\lambda}+F^{\lambda 0}\big)\Big]+\frac{1}{R^{2}}\Big[x_{\mu}x^{\alpha}\big(\partial_{\alpha}F^{\mu\lambda}
+\partial_{\alpha}F^{\lambda\mu}\big)-2x_{\mu}\big(F^{\mu\lambda}+F^{\lambda\mu}\big)\Big]=0.
\end{eqnarray}
Now, we define the four-current as follows

\begin{eqnarray}\label{equation33}
\mu_{0}J^{\lambda}\hspace{-0.2cm}&=&\hspace{-0.2cm}-\partial_{\mu}F^{\lambda\mu}+\frac{1}{R}\Big[\beta x^{\mu}\partial_{\mu}F^{\lambda 0}+\chi x_{\mu}\partial_{0}F^{\lambda\mu}-2\sigma F^{\lambda 0}\Big]\nonumber\\
&&\hspace{1.4cm}-\frac{1}{R^{2}}\Big[\rho x_{\mu}x^{\alpha}\partial_{\alpha}F^{\lambda\mu}-2\varkappa x_{\mu}F^{\lambda\mu}\Big],
\end{eqnarray}
where $\beta$, $\chi$, $\sigma$, $\rho$ and $\varkappa$ are constants where each one of them can take as values only $-1$, $0$ or $1$. To fix them, we impose that the well-known electric-magnetic symmetry

\begin{equation}
\overrightarrow{E} \mapsto c\overrightarrow{B},   \ \ \ \ \ \ \ \ \ \ \ \
   \overrightarrow{B} \rightarrow - \frac{1}{c}\overrightarrow{E},
\end{equation}
remains satisfied in the absence of source. Indeed, by using the definitions of the electric and magnetic fields

\begin{equation}\label{equation34}
E^{i}=cF^{i0}, \hskip35mm B_{i}=\varepsilon_{ijk}F^{jk}/2,
\end{equation}
where $\varepsilon_{ijk}$ is the Levi-Civita antisymmetric tensor $(\varepsilon_{123}=1)$, we can easily deduce that the development of Eqs. (\ref{equation31}) and (\ref{equation33}) gives us $\beta=\chi=\sigma=\rho=\varkappa=1$. With this configuration of values, the exact form of the generalized second group of Maxwell's equations is expressed as follows

\begin{multline}
\partial_{\mu}F^{\mu\lambda}-\frac{1}{R}\Big[x_{\mu}\partial^{\mu}F^{0\lambda}+ x_{\mu}\partial_{0}F^{\mu\lambda}-2 F^{0\lambda}\Big]\\
+\frac{1}{R^{2}}\Big[ x_{\mu}x^{\alpha}\partial_{\alpha}F^{\mu\lambda}-2 x_{\mu}F^{\mu\lambda}\Big]=\mu_{0}J^{\lambda}.
\end{multline} 
Finally, in the explicit form, the exact form of the generalized Maxwell's equations are written as follows

\begin{eqnarray}
\label{equation34a}
\overrightarrow{\nabla}\cdot\overrightarrow{B}+\frac{1}{R}\overrightarrow{r}\cdot\partial_{0}\overrightarrow{B}+\frac{1}{R^{2}}\Big[2\overrightarrow{r}\cdot\overrightarrow{B}-\overrightarrow{r}\cdot x^{\alpha}\partial_{\alpha}\overrightarrow{B}\Big]\hspace{-0.2cm}&=&\hspace{-0.2cm}0,\\
\label{equation34b}
\overrightarrow{\nabla}\cdot\overrightarrow{E}+\frac{1}{R}\overrightarrow{r}\cdot\partial_{0}\overrightarrow{E}+\frac{1}{R^{2}}\Big[2\overrightarrow{r}\cdot\overrightarrow{E}-\overrightarrow{r}\cdot x^{\alpha}\partial_{\alpha}\overrightarrow{E}\Big]\hspace{-0.2cm}&=&\hspace{-0.2cm}\mu_{0}j^{0},\\
\label{equation34c}
   \frac{1}{c}\overrightarrow{\nabla} \wedge \overrightarrow{E} + \partial_{0} \overrightarrow{B}
+ \frac{1}{R}\Big[ \frac{1}{c}\big(\overrightarrow{r} \wedge \partial_{0} \overrightarrow{E}\big)+2\overrightarrow{B}-x^{0} \partial_{0} \overrightarrow{B}\nonumber\\
 - x^{\alpha}\partial_{\alpha} \overrightarrow{B}\Big] +\frac{1}{R^{2}}\Big[\frac{1}{c}\big(2\overrightarrow{r}\wedge\overrightarrow{E}-\overrightarrow{r}\wedge x^{\alpha}\partial_{\alpha}\overrightarrow{E}\big)\nonumber\\
+x^{0}\big(x^{\alpha}\partial_{\alpha}\overrightarrow{B}-2\overrightarrow{B}\big)\Big]\hspace{-0.2cm}&=&\hspace{-0.2cm}\overrightarrow{0},\\
\label{equation34d}
   \overrightarrow{\nabla} \wedge \overrightarrow{B} -\frac{1}{c} \partial_{0} \overrightarrow{E}- \frac{1}{R}\Big[-\overrightarrow{r}\wedge \partial_{0}\overrightarrow{B}+\frac{1}{c}\big(2\overrightarrow{E}-x^{0}\partial_{0} \overrightarrow{E}\nonumber\\
  -x^{\alpha}\partial_{\alpha}\overrightarrow{E}\big)\Big]+\frac{1}{R^{2}}\Big[\big(2\overrightarrow{r}\wedge\overrightarrow{B}-\overrightarrow{r}\wedge x^{\alpha}\partial_{\alpha}\overrightarrow{B}\big)\nonumber\\
+\frac{x^{0}}{c}\big(2\overrightarrow{E}-x^{\alpha}\partial_{\alpha}\overrightarrow{E}\big)\Big]\hspace{-0.2cm}&=&\hspace{-0.2cm}\mu_{0}\overrightarrow{j}.
\end{eqnarray}
To first order in $1/R$, we recover the results obtained in \cite{Takka1}. Before finishing with this second section, let us talk a little about the generalized Lorentz force. In fact, using the generalized Newton's force \cite{Tanimura}, the derivation of $[x^{\mu}, \dot{x}^{\nu}]$ with respect to the parameter $\tau$ gives 

\begin{equation}\label{GLF1}
[x^{\mu}, F^{\nu}]=m[x^{\mu}, \ddot{x}^{\nu}]=m\frac{d}{d\tau}[x^{\mu}, \dot{x}^{\nu}]-m[\dot{x}^{\mu}, \dot{x}^{\nu}].
\end{equation}
By means of equations (\ref{equation7}) and (\ref{equation26}), expression (\ref{GLF1}) reproduces the usual result of special relativity and the one established to first order in the $R$-Minkowski space-time \cite{Takka1}. To second order in $1/R$, the generalized Lorentz force can be thus expressed as 

\begin{multline}\label{GLF6}
\hspace{-0.3cm} F^{\nu}_{(2)}=q\big<F^{\nu\lambda}\dot{x}_{\lambda}\big>_{(2)}+G_{0}^{\nu}(x)
+\frac{1}{R}\left\{-m\big(\dot{x}^{\nu}\dot{x}^{0}+\dot{x}^{0}\dot{x}^{\nu}\big)+q\Big(2\big<F^{\nu\lambda}x_{0}\dot{x}_{\lambda}\big>\right.\\
\left.+\big<F^{\nu 0}x^{\lambda}\dot{x}_{\lambda}\big>+\big<F^{\nu\lambda}x_{\lambda}\dot{x}_{0}\big>\Big)+G_{1}^{\nu}(x)\right\}_{(1)}+\frac{1}{R^{2}}D^{\nu}_{(0)}(x, \dot{x}),
\end{multline}
where $G_{0}^{\nu}(x)$ and $G_{1}^{\nu}(x)$ are integration functions. $D^{\nu}_{(0)}$ is an unknown function of zeroth order in $1/R$ containing the corresponding corrective terms. By looking at expression (\ref{GLF6}), it is notable that the particle mass intervenes in the calculation of the generalized Lorentz force. This new parameter makes that any two particles with the same charge, submitted to the same electromagnetic field, will not necessarily feel the same force if their masses are different. To first order in $1/R$, we can see that for a neutral particle, the aforementioned force is proportional to $m/R$. The same conclusion was found in the $k$-Minkowski space-time \cite{Harikumar2} constituting then a common point between the two deformed Minkowski spaces. Using the iterative method that we have developed in \cite{Takka1}, we can find $F^{\nu}_{(2)}$ after a very long calculation. The same approach may be used to go further in order to find the higher-order terms.

\section{The $R$-Lorentz algebra symmetry}
In this section, we will examine the hypothesis of the existence of Dirac's magnetic monopole by studying the impact that induces the $R$-deformation of the three commutators $[x^{\mu}, x^{\nu}]$, $[x^{\mu}, \dot{x}^{\nu}]$ and $[\dot{x}^{\mu}, \dot{x}^{\nu}]$ on the $R$-Lorentz algebra satisfying the following relations \cite{Foughali-Bouda1}

\begin{align}
\label{Rlalg1}
\big[x^{\mu}, \textbf{J}^{\nu\lambda}\big] & = -i\hbar\big(\eta^{\mu\lambda}x^{\nu}-\eta^{\mu\nu}x^{\lambda}\big) +
                                       \frac{i\hbar}{R}x^{\mu}\big(\eta^{0\lambda}x^{\nu}-\eta^{0\nu}x^{\lambda}\big), \\
\label{Rlalg2}
\big[p^{\mu}, \textbf{J}^{\nu\lambda}\big] & =  i\hbar\big(\eta^{\nu\mu}p^{\lambda}-\eta^{\lambda\mu}p^{\nu}\big)+
                                       \frac{i\hbar}{R}p^{\mu}\big(\eta^{\nu 0}x^{\lambda}-\eta^{\lambda 0}x^{\nu}\big),\\
\label{Rlalg3} 
\big[\textbf{J}^{\mu\nu}, \textbf{J}^{\lambda\sigma}\big] & = i\hbar \big(\eta^{\nu\lambda}\textbf{J}^{\mu\sigma}-
                                                   \eta^{\nu\sigma}\textbf{J}^{\mu\lambda}
                                              -\eta^{\mu\lambda}\textbf{J}^{\nu\sigma}+\eta^{\mu\sigma}\textbf{J}^{\nu\lambda}\big),
\end{align}
where $\textbf{J}^{\mu \nu }$ is the total angular momentum expressed as

\begin{equation}\label{angular momentum}
\textbf{J}^{\mu \nu }=x^{\mu }p^{\nu }-x^{\nu }p^{\mu }.
\end{equation}
As a start point, we come back to the expression of the four-dimensional momentum valid in the absence of the electromagnetic field   where \cite{Bouda-Foughali}

\begin{equation}\label{RLS1}
p^{\mu}=m\Big(1-\frac{x^{0}}{R}\Big)^{-2}\dot{x}^{\mu}+\frac{m}{R}\Big(1-\frac{x^{0}}{R}\Big)^{-3}x^{\mu}\dot{x}^{0}.
\end{equation}
When there is no field, we can use Eq. (\ref{RLS1}) to rewrite relation (\ref{angular momentum}) by means of coordinates and velocities

\begin{equation}\label{RLS2}
J^{\mu\nu}= m\Big(1-\frac{x^{0}}{R}\Big)^{-2}\big(x^{\mu}\dot{x}^{\nu} - x^{\nu}\dot{x}^{\mu}\big).
\end{equation}
Taking into account Eqs. (\ref{equation2}), (\ref{equation7}), (\ref{equationaj1}), (\ref{equation13r}) and (\ref{equationsym6}),  after symmetrization, we arrive to
\begin{equation}\label{RLS3}
J^{\mu\nu}= m\Big(1-\frac{x^{0}}{R}\Big)^{-2}\big(x^{\mu}\dot{x}^{\nu}-x^{\nu}\dot{x}^{\mu}\big)-\frac{i\hbar}{R}\big(\eta^{\mu 0}x^{\nu}-\eta^{\nu 0}x^{\mu}\big).
\end{equation}
In the absence of the field, relations (\ref{equation2}), (\ref{equation7}), (\ref{equation22}) and (\ref{RLS3}) lead  to the following $R$-Lorentz algebra

\begin{align}
\label{RLS5abs}
\big[x^{\mu}, J^{\nu\lambda}\big]&=i\hbar\big(\eta^{\mu\nu}x^{\lambda}-\eta^{\mu\lambda}x^{\nu}\big)
+\frac{i\hbar}{R}\big(\eta^{0\lambda}x^{\nu}-\eta^{0\nu}x^{\lambda}\big)x^{\mu}, \\
\label{RLS6abs}
\big[\dot{x}^{\mu}, J^{\nu\lambda}\big]&= i\hbar \big(\eta^{\mu\nu} \dot{x}^{\lambda} - \eta^{\mu\lambda} \dot{x}^{\nu}\big)-\frac{i\hbar}{R} \Big\{x^{\mu}\big(\eta^{0\nu} \dot{x}^{\lambda}- \eta^{0\lambda} \dot{x}^{\nu} \big)\nonumber\\
&\hspace{0.4cm}+\big(\eta^{0\nu} x^{\lambda}-\eta^{0\lambda} x^{\nu} \big)\dot{x}^{\mu}\Big\}- \frac{\hbar^{2}}{mR}\Big(1-\frac{x^{0}}{R}\Big)^{2}\Big\{\big(\eta^{0\lambda}\eta^{\mu\nu} \nonumber\\
&\hspace{0.4cm}- \eta^{0\nu} \eta^{\mu\lambda} \big)+\frac{1}{R}\Big(\eta^{\mu 0}-\frac{x^{\mu}}{R}\Big)\big(\eta^{\nu 0}x^{\lambda}- \eta^{\lambda 0} x^{\nu} \big)\Big\}\\
\label{RLS7abs}
\big[J^{\mu\nu}, J^{\lambda\sigma}\big]&=i\hbar\big(\eta^{\nu\lambda}J^{\mu\sigma}
-\eta^{\nu\sigma}J^{\mu\lambda}- \eta^{\mu\lambda}J^{\nu\sigma} + \eta^{\mu\sigma}J^{\nu\lambda}\big).
\end{align}
We observe that the two commutators (\ref{RLS5abs}) and (\ref{RLS7abs}) have the same forms as in (\ref{Rlalg1}) and (\ref{Rlalg3}) respectively. Relation (\ref{RLS6abs}) can be derived from (\ref{Rlalg2}). Now, if we note by $M^{\mu\nu}$ the contribution of the electromagnetic field to the total angular momentum $\textbf{J}^{\mu\nu}$, expression (\ref{RLS3}) becomes therefore

\begin{eqnarray}\label{RLS4}
\textbf{J}^{\mu\nu} \hspace{-0.2cm}&=&\hspace{-0.2cm}J^{\mu\nu}+M^{\mu\nu}\nonumber\\
                    \hspace{-0.2cm}&=&\hspace{-0.2cm} m\Big(1-\frac{x^{0}}{R}\Big)^{-2}\big(x^{\mu}\dot{x}^{\nu}-x^{\nu}\dot{x}^{\mu}\big)-\frac{i\hbar}{R}\big(\eta^{\mu 0}x^{\nu}-\eta^{\nu 0}x^{\mu}\big)+M^{\mu\nu}.
\end{eqnarray}
From this last expression, by taking into account relations (\ref{equation2}), (\ref{equation7}) and (\ref{equation26}),  we find the following $R$-Lorentz algebra

\begin{align}
\label{RLS5}
\hspace{-0.5cm}\big[x^{\mu}, \textbf{J}^{\nu\lambda}\big]&=i\hbar\big(\eta^{\mu\nu}x^{\lambda}-\eta^{\mu\lambda}x^{\nu}\big)
+\frac{i\hbar}{R}\big(\eta^{0\lambda}x^{\nu}-\eta^{0\nu}x^{\lambda}\big)x^{\mu}+\big[x^{\mu},M^{\nu\lambda}\big], \\ \nonumber\\
\label{RLS6}
\hspace{-0.5cm}\big[\dot{x}^{\mu},\textbf{J}^{\nu\lambda}\big]&= i\hbar \big(\eta^{\mu\nu} \dot{x}^{\lambda} - \eta^{\mu\lambda} \dot{x}^{\nu}\big)-\frac{i\hbar}{R} \Big\{x^{\mu}\big(\eta^{0\nu} \dot{x}^{\lambda}- \eta^{0\lambda} \dot{x}^{\nu} \big)+\big(\eta^{0\nu} x^{\lambda}-\nonumber\\
&\hspace{0.4cm}\eta^{0\lambda} x^{\nu} \big)\dot{x}^{\mu}\Big\}- \frac{\hbar^{2}}{mR}\Big(1-\frac{x^{0}}{R}\Big)^{2}\Big\{\big(\eta^{0\lambda}\eta^{\mu\nu}- \eta^{0\nu} \eta^{\mu\lambda} \big)+\frac{1}{R}\big(\eta^{\mu 0}-\nonumber\\
&\hspace{0.4cm}\frac{x^{\mu}}{R}\big)\big(\eta^{\nu 0}x^{\lambda}- \eta^{\lambda 0} x^{\nu} \big)\Big\}-\frac{i\hbar q}{m} \Big(1-\frac{x^{0}}{R}\Big)^{-2}\Big\{\big(x^{\nu}F^{\mu\lambda}- x^{\lambda}F^{\mu\nu}\big)\Big\}\nonumber\\
&\hspace{0.4cm}+\big[\dot{x}^{\mu},M^{\nu\lambda}\big],\\ \nonumber\\
\label{RLS7}
\hspace{-0.5cm}\big[\textbf{J}^{\mu\nu}, \textbf{J}^{\lambda\sigma}\big]&=i\hbar\big(\eta^{\nu\lambda}\textbf{J}^{\mu\sigma}
-\eta^{\nu\sigma}\textbf{J}^{\mu\lambda}- \eta^{\mu\lambda}\textbf{J}^{\nu\sigma} + \eta^{\mu\sigma}\textbf{J}^{\nu\lambda}\big),\nonumber\\
&=i\hbar\big(\eta^{\nu\lambda}J^{\mu\sigma}-\eta^{\nu\sigma}J^{\mu\lambda}- \eta^{\mu\lambda}J^{\nu\sigma} + \eta^{\mu\sigma}J^{\nu\lambda}\big)\nonumber\\
&\hspace{0.4cm}+i\hbar q \Big(1-\frac{x^{0}}{R}\Big)^{-4}\Big\{\big(x^{\mu}x^{\sigma}F^{\nu\lambda}- x^{\mu}x^{\lambda}F^{\nu\sigma} - x^{\nu}x^{\sigma}F^{\mu\lambda}\nonumber\\
&\hspace{0.4cm}+x^{\nu}x^{\lambda}F^{\mu\sigma}\big)\Big\}+\big[J^{\mu\nu}, M^{\lambda\sigma}\big]+\big[M^{\mu\nu}, J^{\lambda\sigma}\big]+ \big[M^{\mu\nu}, M^{\lambda\sigma}\big].
\end{align}
To seek the conditions under which the above algebra remains invariant in the presence of electromagnetic field, we should determine the suitable form of $M^{\mu\nu}$. In other words, the restoration of the $R$-Lorentz algebra constituted by relations (\ref{Rlalg1}), (\ref{Rlalg2}) and (\ref{Rlalg3}) and represented in the absence of the electromagnetic field by relations (\ref{RLS5abs}), (\ref{RLS6abs}) and (\ref{RLS7abs}) respectively, requires the following three conditions:\\

\begin{align}
\label{RLS8}
&\big[x^{\mu}, M^{\nu\lambda}\big] = 0,\\
\label{RLS9}
&\big[\dot{x}^{\mu}, M^{\nu\lambda}\big]=\frac{i\hbar q}{m}\Big(1-\frac{x^{0}}{R}\Big)^{-2}\big(x^{\nu}F^{\mu\lambda}-x^{\lambda}F^{\mu\nu}\big),\\
\label{RLS10}
& i\hbar \big(\eta^{\nu\lambda}M^{\mu\sigma} - \eta^{\nu\sigma}M^{\mu\lambda} - \eta^{\mu\lambda}M^{\nu\sigma} + \eta^{\mu\sigma}M^{\nu\lambda}\big)\nonumber\\
&=i\hbar q\Big(1-\frac{x^{0}}{R}\Big)^{-4}\big(x^{\mu}x^{\sigma}F^{\nu\lambda}- x^{\mu}x^{\lambda}F^{\nu\sigma} - x^{\nu}x^{\sigma}F^{\mu\lambda}+ x^{\nu}x^{\lambda}F^{\mu\sigma}\big)\nonumber\\
& \hspace{0.4cm}+ \big[J^{\mu\nu}, M^{\lambda\sigma}\big] + \big[M^{\mu\nu}, J^{\lambda\sigma}\big] + \big[M^{\mu\nu}, M^{\lambda\sigma}\big].
\end{align}
Relation (\ref{RLS8}) means that $M^{\nu\lambda}$ depends on position but not on velocity. Using both expressions (\ref{RLS8}) and (\ref{RLS9}), we can check that Eqs. (\ref{equation2}) and (\ref{RLS3}) allow us to have

\begin{align}\label{RLS11}
\big[J^{\mu\nu}, M^{\lambda\sigma}\big]&=\Big[m\Big(1-\frac{x^{0}}{R}\Big)^{-2}\big(x^{\mu}\dot{x}^{\nu}-x^{\nu}\dot{x}^{\mu}\big), M^{\lambda\sigma}\Big],\nonumber\\
                               &=m\Big(1-\frac{x^{0}}{R}\Big)^{-2}\Big\{x^{\mu}\big[\dot{x}^{\nu},M^{\lambda\sigma}\big] -x^{\nu}\big[\dot{x}^{\mu}, M^{\lambda\sigma}\big]\Big\}\nonumber\\
                               &=i\hbar q \big(1-\frac{x^{0}}{R}\big)^{-4}\Big\{x^{\mu}x^{\lambda}F^{\nu\sigma}- x^{\mu}x^{\sigma}F^{\nu\lambda} - x^{\nu}x^{\lambda}F^{\mu\sigma}\nonumber\\
                               &\hspace{6.2cm} + x^{\nu}x^{\sigma}F^{\mu\lambda}\Big\}.
\end{align}
In the same way, we find

\begin{eqnarray}\label{RLS12}
\big[M^{\mu\nu}, J^{\lambda\sigma}\big]\hspace{-0.2cm}&=&\hspace{-0.2cm}i\hbar q \Big(1-\frac{x^{0}}{R}\Big)^{-4}\Big\{x^{\mu}x^{\lambda}F^{\nu\sigma}- x^{\mu}x^{\sigma}F^{\nu\lambda} - x^{\nu}x^{\lambda}F^{\mu\sigma}\nonumber\\
&&\hspace{5.9cm} + x^{\nu}x^{\sigma}F^{\mu\lambda}\Big\}.
\end{eqnarray}
Because $M^{\mu\nu}=M^{\mu\nu}(x)$ which implies that $\big[M^{\mu\nu}, M^{\lambda\sigma}\big]=0$, the substitution of Eqs. (\ref{RLS11}) and (\ref{RLS12}) in (\ref{RLS10}) gives

\begin{multline}\label{RLS13}
\big(\eta^{\nu\lambda}M^{\mu\sigma} - \eta^{\nu\sigma}M^{\mu\lambda} - \eta^{\mu\lambda}M^{\nu\sigma} + \eta^{\mu\sigma}M^{\nu\lambda}\big)=\\
q\Big(1-\frac{x^{0}}{R}\Big)^{-4}\big(x^{\mu}x^{\lambda}F^{\nu\sigma}-x^{\mu}x^{\sigma}F^{\nu\lambda} - x^{\nu}x^{\lambda}F^{\mu\sigma}+x^{\nu}x^{\sigma}F^{\mu\lambda}\big).
\end{multline}
Focusing only on the spatial part of expression (\ref{RLS13}), we replace $(\mu, \nu, \lambda, \sigma)$ by $(i,j,k,l)$ respectively. After the contraction of the resulting equation by $\eta_{il}$, we find

\begin{equation}\label{RLS14}
M^{jk}=q\Big(1-\frac{x^{0}}{R}\Big)^{-4}\big(x^{k}x_{l}F^{jl}-x_{l}x^{l}F^{jk}+x^{j}x_{l}F^{lk}\big).
\end{equation}
Using the definition according to which the magnetic angular momentum \cite{Berard2} takes this form

\begin{equation}\label{RLS15}
M_{i} \equiv \frac{1}{2} \varepsilon_{ijk} M^{jk},
\end{equation}
Eq.(\ref{RLS14}) gives

\begin{equation}\label{RLS16}
M_{i}=\frac{1}{2}\varepsilon_{ijk}q\Big(1-\frac{x^{0}}{R}\Big)^{-4}\big(x^{k}x_{l}F^{jl}-x_{l}x^{l}F^{jk}+x^{j}x_{l}F^{lk}\big).
\end{equation}
Taking into account the following relation

\begin{equation}\label{RLS17}
F^{ij} = \varepsilon^{kij} B^{k},
\end{equation}
we can prove that Eq. (\ref{RLS16}) reduces to

\begin{equation}\label{RLS18}
M^{i}=q\Big(1-\frac{x^{0}}{R}\Big)^{-4}(x^{j}B^{j})x^{i},
\end{equation}
which can be written in vector form as

\begin{equation}\label{RLS19}
\overrightarrow{M}=q\Big(1-\frac{x^{0}}{R}\Big)^{-4}\big(\overrightarrow{r}\cdot\overrightarrow{B}\big)\overrightarrow{r}.
\end{equation}
Because $M^{jk}$ depends only on coordinates, Eqs. (\ref{commutatif}) and (\ref{equation7}) let write

\begin{align}\label{RLS20}
\big[\dot{x}^{i}, M^{jk}\big] & = [\dot{x}^{i}, x^{\mu}] \frac{\partial M^{jk}}{\partial x^{\mu}} \nonumber \\
                      & = \frac{i\hbar}{m}\Big(1-\frac{x^{0}}{R}\Big)^{2}\left\{\frac{\partial M^{jk}}{\partial x_{i}}
                          - \frac{x^{i}}{R}\Big(1-\frac{x^{0}}{R}\Big)\frac{\partial M^{jk}}{\partial x_{0}}+\frac{x^{n}x^{i}}{R^{2}}\frac{\partial M^{jk}}{\partial x^{n}}\right\}.
\end{align}
After the identification of Eq. (\ref{RLS20}) with the spatial part of Eq. (\ref{RLS9}), we get

\begin{equation}\label{RLS21}
\frac{\partial M^{jk}}{\partial x_{i}} - \frac{x^{i}}{R}\Big(1-\frac{x^{0}}{R}\Big)\frac{\partial M^{jk}}{\partial x_{0}}+\frac{x^{n}x^{i}}{R^{2}}\frac{\partial M^{jk}}{\partial x^{n}}=
q\Big(1-\frac{x^{0}}{R}\Big)^{-4}\big(x^{j}F^{ik}-x^{k}F^{ij}\big).
\end{equation}
Multiplying Eq. (\ref{RLS21}) by $\varepsilon^{ljk}/2$ and exploiting Eqs. (\ref{RLS15}) and (\ref{RLS17}), we find 

\begin{equation}\label{RLS21a}
\frac{\partial M^{l}}{\partial x_{i}} - \frac{x^{i}}{R}\Big(1-\frac{x^{0}}{R}\Big)\frac{\partial M^{l}}{\partial x_{0}}+\frac{x^{i}x^{n}}{R^{2}}\frac{\partial M^{l}}{\partial x^{n}}=
q\Big(1-\frac{x^{0}}{R}\Big)^{-4}\big(B^{l}x^{i}-\overrightarrow{r}\cdot\overrightarrow{B}\delta^{li}\big).
\end{equation}
In the stationary case where $\overrightarrow{B}$ does not depend on $x^{0}$, the developement of Eq. (\ref{RLS21a}) by means of (\ref{RLS18}) gives rise to the following relation

\begin{equation}\label{RLS22}
x^{i}B^{l} + B^{i}x^{l}-x^{j}\frac{\partial B^{j}}{\partial x_{i}}x^{l}+\frac{x^{i}x^{j}x^{l}}{R^{2}}\Big\{2B^{j}-x^{n}\frac{\partial B^{j}}{\partial x^{n}}\Big\}=0.
\end{equation} 
To zeroth order in $1/R$, relation (\ref{RLS22}) reduces to the usual expression of the special relativity \cite{Berard2} 

\begin{equation}\label{RLS22order0}
x^{i}B^{l}_{0} + B^{i}_{0}x^{l}-x^{j}\frac{\partial B^{j}_{0}}{\partial x_{i}}x^{l}=0,
\end{equation}
admitting as solution 

\begin{equation}\label{RLS25}
B^{i}_{0} = \frac{g_{0}}{4\pi}  \frac{x^{i}}{r^{3}},
\end{equation}
where $r=\left(x^{j}x^{j}\right)^{1/2}$ and $g_{0}$ describes the elementary magnetic charge. To first order in $1/R$, the result remains unchanged \cite{Takka1}. Now to go further, we have opted for the use of the perturbative treatment. To first order in $1/R^{2}$, let seek the solution in the form of

\begin{equation}\label{RLS26}
B^{i}=B^{i}_{0}+\frac{B^{i}_{2}}{R^{2}}=\frac{g_{0}}{4\pi}  \frac{x^{i}}{r^{3}}+\frac{B^{i}_{2}}{R^{2}}.
\end{equation}
By replacing the last expression in Eq. (\ref{RLS22}) and taking into account Eq. (\ref{RLS22order0}), we get

\begin{equation}\label{RLS28}
x^{i}B^{l}_{2} + B^{i}_{2}x^{l}-x^{j}\frac{\partial B^{j}_{2}}{\partial x_{i}}x^{l}+\frac{g_{0}}{\pi}\frac{x^{i}x^{l}}{r}=0.
\end{equation}
The general solution of the last equations is written as the sum of the homogeneous and the particular solutions given by

\begin{equation}\label{RLS29}
B^{i}_{2}=\frac{g_{2}}{4\pi}  \frac{x^{i}}{r^{3}}+\frac{\gamma_{2} g_{0}}{\pi}\frac{x^{i}}{r},\hspace{0.1cm}\gamma_{2}\in\mathbb{R^{*}}.
\end{equation}
For the moment, $g_{2}$ is just an integration constant. After the substitution of Eq. (\ref{RLS29}) in (\ref{RLS28}), we find 

\begin{equation}\label{RLS31sol}
\gamma_{2}=-\frac{1}{2},
\end{equation}
and then expression (\ref{RLS26}) can be explicited as

\begin{equation}\label{RLS32}
B^{i}=\frac{g_{0}}{4\pi}\frac{x^{i}}{r^{3}}+\frac{1}{R^{2}}\left(\frac{g_{2}}{4\pi}  \frac{x^{i}}{r^{3}}-\frac{g_{0}}{2\pi}\frac{x^{i}}{r}\right).
\end{equation}
Next, to second order in $1/R^{2}$, we write $B^{i}$ as follows

\begin{equation}\label{RLS33}
B^{i}=\frac{g_{0}}{4\pi}\frac{x^{i}}{r^{3}}+\frac{1}{R^{2}}\left(\frac{g_{2}}{4\pi}  \frac{x^{i}}{r^{3}}-\frac{g_{0}}{2\pi}\frac{x^{i}}{r}\right)+\frac{B^{i}_{4}}{R^{4}}.
\end{equation}
Taking into account Eqs. (\ref{RLS22order0}) and (\ref{RLS28}), the replacement of Eq. (\ref{RLS33}) in (\ref{RLS22}) yields

\begin{equation}\label{RLS35}
x^{i}B^{l}_{4} + B^{i}_{4}x^{l}-x^{j}\frac{\partial B^{j}_{4}}{\partial x_{i}}x^{l}+\frac{g_{2}}{\pi}\frac{x^{i}x^{l}}{r}-\frac{g_{0}}{\pi}x^{i}x^{l}r=0.
\end{equation}
Seeking a solution in the form of

\begin{equation}\label{RLS36}
B^{i}_{4}=\frac{g_{4}}{4\pi}  \frac{x^{i}}{r^{3}}-\frac{g_{2}}{2\pi}\frac{x^{i}}{r}+\frac{\gamma_{4} g_{0}}{\pi}x^{i}r,\hspace{0.1cm}\gamma_{4}\in\mathbb{R^{*}},
\end{equation}
where $g_{4}$ is an integration constant, the substitution of Eq. (\ref{RLS36}) in (\ref{RLS35}) gives $\gamma_{4}=1/4$ and then

\begin{equation}\label{RLS37}
B^{i}_{4} = \frac{g_{4}}{4\pi}  \frac{x^{i}}{r^{3}}-\frac{g_{2}}{2\pi}\frac{x^{i}}{r}+\frac{g_{0}}{4\pi}x^{i}r,
\end{equation}
and therefore, expression (\ref{RLS33}) becomes

\begin{equation}\label{RLS38}
B^{i}=\frac{g_{0}}{4\pi}\frac{x^{i}}{r^{3}}+\frac{1}{R^{2}}\left(\frac{g_{2}}{4\pi}  \frac{x^{i}}{r^{3}}-\frac{g_{0}}{2\pi}\frac{x^{i}}{r}\right)+\frac{1}{R^{4}}\left(\frac{g_{4}}{4\pi}  \frac{x^{i}}{r^{3}}-\frac{g_{2}}{2\pi}\frac{x^{i}}{r}+\frac{g_{0}}{4\pi}x^{i}r\right).
\end{equation}
Similarly, to third order in $1/R^{2}$, we have

\begin{equation}\label{RLS39}
B^{i}=\frac{g_{0}}{4\pi}\frac{x^{i}}{r^{3}}+\frac{1}{R^{2}}\left(\frac{g_{2}}{4\pi}  \frac{x^{i}}{r^{3}}-\frac{g_{0}}{2\pi}\frac{x^{i}}{r}\right)+\frac{1}{R^{4}}\left(\frac{g_{4}}{4\pi}  \frac{x^{i}}{r^{3}}-\frac{g_{2}}{2\pi}\frac{x^{i}}{r}+\frac{g_{0}}{4\pi}x^{i}r\right)+\frac{B^{i}_{6}}{R^{6}}.
\end{equation}
The use of (\ref{RLS22order0}), (\ref{RLS28}) and (\ref{RLS35}) after the replacement of Eq. (\ref{RLS39}) in (\ref{RLS22}) gives

\begin{equation}\label{RLS41}
x^{i}B^{l}_{6} + B^{i}_{6}x^{l}-x^{j}\frac{\partial B^{j}_{6}}{\partial x_{i}}x^{l}+\frac{g_{4}}{\pi}\frac{x^{i}x^{l}}{r}-\frac{g_{2}}{\pi}x^{i}x^{l}r=0,
\end{equation}
which has an identical form as that of the second order approximation (\ref{RLS35}) where the indices $(4, 2, 0 )$ are replaced by $(6, 4, 2 )$ respectively. As a consequence, the third order corrective term in $1/R^{2}$ is expressed as in $(\ref{RLS37})$:

\begin{equation}\label{RLS42}
B^{i}_{6} = \frac{g_{6}}{4\pi}  \frac{x^{i}}{r^{3}}-\frac{g_{4}}{2\pi}\frac{x^{i}}{r}+\frac{g_{2}}{4\pi}x^{i}r,
\end{equation}
where $g_{6}$ is another integration constant. Now, without doing any calculation, we deduce that the higher order corrective terms admit the same form as that of $B^{i}_{4}$ but distinguished by the corresponding integration constants. Mathematically, the exact form of the generalized Dirac's magnetic monopole can then be written as

\begin{multline}\label{RLS43}
\hspace{-0.3cm} B^{i}=\lim\limits_{n \rightarrow +\infty}\left[\frac{g_{0}}{4\pi}\frac{x^{i}}{r^{3}}+\frac{1}{R^{2}}\left(\frac{g_{2}}{4\pi}  \frac{x^{i}}{r^{3}}-\frac{g_{0}}{2\pi}\frac{x^{i}}{r}\right)+\frac{1}{R^{4}}\left(\frac{g_{4}}{4\pi}  \frac{x^{i}}{r^{3}}-\frac{g_{2}}{2\pi}\frac{x^{i}}{r}+\frac{g_{0}}{4\pi}x^{i}r\right)\right.\\ \left.+...+\frac{1}{R^{2n}}\left(\frac{g_{2n}}{4\pi}  \frac{x^{i}}{r^{3}}-\frac{g_{2n-2}}{2\pi}\frac{x^{i}}{r}+\frac{g_{2n-4}}{4\pi}x^{i}r\right)\right],\hspace{0.2cm}n\in\{\mathbb{N}-\{0,1\}\},
\end{multline}
which is equivalent to
 
\begin{equation}\label{RLS44}
B^{i}=\lim\limits_{n \rightarrow +\infty}\frac{1}{4\pi}\left( g_{0}+\frac{g_{2}}{R^{2}}+\frac{g_{4}}{R^{4}}+\frac{g_{6}}{R^{6}}+...+\frac{g_{2n}}{R^{2n}}\right)\Big(1-\frac{r^{2}}{R^{2}}\Big)^{2}\frac{x^{i}}{r^{3}},
\end{equation}
or finally to

\begin{equation}\label{RLS45}
B^{i}=\frac{g_{e}}{4\pi}\Big(1-\frac{r^{2}}{R^{2}}\Big)^{2}\frac{x^{i}}{r^{3}},
\end{equation}
where

\begin{equation}\label{RLS46}
g_{e}\equiv\lim\limits_{n \rightarrow +\infty}\left( g_{0}+\frac{g_{2}}{R^{2}}+\frac{g_{4}}{R^{4}}+\frac{g_{6}}{R^{6}}+...+\frac{g_{2n}}{R^{2n}}\right).
\end{equation}
Here $g_{e}$ represents a new characteristic quantity having the dimension of the magnetic charge and which can in general depend on the radius of the Universe.  By observing (\ref{RLS46}), we can see that in the limit case where $R$ goes to the infinity, this latter reduces to the elementary magnetic charge of special relativity. In the general case, to determine the quantity of the magnetic charge $ q_ {m} $ that could contain a volume delimited by a closed spherical surface in our context, we calculate the flux of the magnetic field through this same surface to have

\begin{equation}\label{RLS49}
q_{m}(r, R)=\oint \vec{B}\cdot d\vec{s}=g_{e}\Big(1-\frac{r^{2}}{R^{2}}\Big)^{2}.
\end{equation}
Schematically, the last result implies that

\begin{eqnarray}\label{RLS50}
   \left\{
     \begin{array}{l}
      \text{if}\hspace{0.2cm}\frac{r}{R}\longrightarrow 0, \hspace{0.5cm}q_{m}(r, R)\longrightarrow g_{0}, \\ \\
     \text{if}\hspace{0.2cm}\frac{r}{R}\longrightarrow 1, \hspace{0.5cm}q_{m}(r, R)\longrightarrow 0.
     \end{array}
  \right.
\end{eqnarray}
By analyzing the above result, we deduce that the magnetic charge could exist locally but its charge is still neutralized in the  Universe . Finally, the replacement of Eq. (\ref{RLS45}) in (\ref{RLS18}) gives the following exact expression of the magnetic angular momentum 

\begin{equation}\label{RLS51}
\overrightarrow{M}=\frac{qg_{e}}{4\pi}\Big(1-\frac{r^{2}}{R^{2}}\Big)^{2}\left(1-\frac{x^{0}}{R}\right)^{-4}\frac{\overrightarrow{r}}{r}.
\end{equation}
For $r=R$ or $x^{0}=R$, it is obvious that $\overrightarrow{M}=\overrightarrow{0}$. It is also easy to check that the first approximation obtained in \cite{Takka1} is recovered.

\newpage
\section{Conclusion}

Taking as a starting point, the equivalence between the minimal coupling prescription and Feynman's approach highlighted in the first paper \cite{Dyson} and confirmed in further works, we found possible to construct a new version in order to generalize electromagnetism in the noncommutative formalism. This approach is directly conditioned by the knowledge of phase space algebra and the explicit form of the four-dimensional momentum valid in the absence of the electromagnetic field. By proceeding in this manner, we have found the exact form of the generalized Maxwell's equations in the context of Fock's nonlinear relativity. As in the $k$-Minkowski space-time \cite{Harikumar2}, the generalized Lorentz force acting on a moving charged particle under the influence of an electromagnetic field depends on the mass of this latter. This new parameter acts so that any two particles with the same
charge, submitted to the same electromagnetic field, will not feel necessarily
the same force if their masses are different. At the end, we have restored the exact form of the $R$-Lorentz algebra symmetry and used the perturbative treatment to resolve the spatial part of the resulting equation. After calculation, we have established the exact form of the generalized Dirac's magnetic monopole. By calculating the quantity of the magnetic charge  that could contain a volume delimited by a closed spherical surface, we have found that the Universe could locally contain the magnetic charge but in its totality it is still neutral. Finally, we emphasize that unlike the previous works where Dirac's magnetic monopole was found without imposing the non-dependence on time despite its similarity with the usual electrostatic field, in our case, the analogy is more symmetric.


\newpage

\begin{thebibliography}{99}
\bibitem{Dyson} F. J. Dyson, Am. J. Phys. \textbf{58}, 209 (1990)
\bibitem{Tanimura} S. Tanimura, Annals Phys. \textbf{220}, 229 (1992), arXiv:hep-th/9306066
\bibitem{Berard1} A. Bérard, Y. Grandati and H. Mohrbach, J. Math. Phys. \textbf{40}, 3732 (1999)
\bibitem{Berard2} A. Bérard, Y. Grandati and H. Mohrbach, Phys. Lett. A \textbf{254}, 133 (1999)
\bibitem{Berard3} A. Bérard and H. Mohrbach, Int. J. Theor. Phys. \textbf{39}, 2623 (2000)
\bibitem{Harikumar1} E. Harikumar, Europhys. Lett. \textbf{90}, 21001 (2010), arXiv:1002.3202
\bibitem{Harikumar2} E. Harikumar, T. Juri\'{c} and S. Meljanac, Phys. Rev. D \textbf{84}, 085020 (2011), arXiv:1107.3936
\bibitem{Montesinos1} M. Montesinos, A. Perez-Lorenzana, Int. J. Theor. Phys. \textbf{38} (1999)
\bibitem{Takka1} N. Takka, A. Bouda and T. Foughali, Can. J. Phys. \textbf{95}, 987 (2017)
\bibitem{Bouda-Foughali} A. Bouda and T. Foughali, Mod. Phys. Lett. A \textbf{27}, 1250036 (2012), arXiv:1204.6397
\bibitem{Ghosh} S. Ghosh and P. Pal, Phys. Rev. D \textbf{75}, 105021 (2007), arXiv:hep-th/0702159
\bibitem{Amelino1} G. Amelino Camelia, Nature \textbf{418}, 34 (2002), arXiv:gr-qc/0207049
\bibitem{Amelino2} G. Amelino Camelia, Phys. Lett. B \textbf{510}, 255 (2001), arXiv:hep-th/0012238
\bibitem{Magueijo1} J. Magueijo and L. Smolin, Phys. Rev. Lett. \textbf{88}, 190403 (2002), arXiv:hep-th/0112090
\bibitem{Magueijo2} J. Magueijo and L. Smolin, Phys. Rev. D \textbf{67}, 044017 (2003), arXiv:gr-qc/0207085
\bibitem{Albrecht-Magueijo} A. Albrecht and J. Magueijo, Phys. Rev. D \textbf{59}, 043516 (1999), arXiv:astro-ph/9811018
\bibitem{Fock} V. Fock, The Theory of Space, Time and Gravitation, Pergamon Press,
Oxford, London, New York, Paris (1964)
\bibitem{Foughali-Bouda1} T. Foughali and A. Bouda, Can. J. Phys. \textbf{93}, 734 (2015), arXiv:1605.01943 
\bibitem{Foughali-Bouda2} T. Foughali and A. Bouda, Int. J. Theor. Phys. \textbf{55}, 2247 (2016), arXiv:1605.04080
\end{thebibliography}
\end{document}